\documentclass[preprint,superscriptaddress,preprintnumbers,amsmath,amssymb]{revtex4}
\usepackage{graphicx}
\usepackage{amsmath}
\usepackage{gensymb}

\begin{document}
\title{Dynamic relaxation oscillations in a nonlinearly driven quartz crystal}
\author{S. Houri}
\affiliation{Kavli Institute of Nanoscience, Delft University of Technology, Lorentzweg 1, 2628 CJ Delft, The Netherlands}
\author{M. J. Geuze}
\affiliation{Kavli Institute of Nanoscience, Delft University of Technology, Lorentzweg 1, 2628 CJ Delft, The Netherlands}
\author{W. J. Venstra}\email{w.j.venstra@tudelft.nl}
\affiliation{Kavli Institute of Nanoscience, Delft University of Technology, Lorentzweg 1, 2628 CJ Delft, The Netherlands}
\affiliation{Quantified Air, Lorentzweg 1, 2628 CJ Delft, The Netherlands}
\date{\today}

\begin{abstract} 
	We demonstrate thermo-mechanical relaxation oscillations in a strongly driven quartz crystal. Dynamic bifurcation leads to two stable oscillation states with a distinct electrical impedance. Slow Joule-heating, which shifts the susceptibility of the crystal, provides a feedback that leads to thermally-induced oscillations, in which the amplitude of the crystal is modulated by a relaxation cycle. The frequency of the relaxation cycle is roughly a million times lower than the resonance frequency of the crystal, and it can be adjusted by the detuning from the critical point for dynamic bifurcation. The experimental observations are reproduced by a simple model that takes into account the slow dynamics of the system. 
\end{abstract}

\maketitle
\indent\indent Harmonic oscillators, such as quartz crystals, are used for timing and sensing purposes, and constitute an indispensable part of modern electronic devices. Besides harmonic oscillators, other types of oscillators such as the relaxation oscillator occur frequently~\cite{jenkins13}. Where harmonic oscillators conserve energy --- they exchange kinetic with potential energy, as in a mass-spring system-- a relaxation oscillator repetitively dissipates energy. It's dynamic behaviour can be described by coupled first-order differential equations, with non-oscillatory steady state limits. The period of a relaxation oscillator is determined by energy decay rates, which result, for example, from viscous friction or resistive dissipation.\\
\indent\indent\ Canonical examples of relaxation oscillators include the electronic flip-flop~\cite{abraham19}, the mammal heart muscle~\cite{vdpol28}, and the Pearson-Anson neon-lamp oscillator~\cite{pearson21}. Mechanical relaxation oscillators have been observed recently in a micro-electromechanical impact device~\cite{bienstman98}, a field-emitting carbon nanotube~\cite{lazarus10}, and in a system of coalescing nanofluidic droplets~\cite{zettl05}. 
Here, we report a \emph{dynamic} relaxation oscillator, with oscillatory steady states, in which the oscillation amplitude of a nonlinear resonator is modulated by a relaxation cycle. The period of the relaxation cycle is determined by the thermal time constant and the mechanical ring-down of the device.\\
\begin{figure}[bh]
\includegraphics[width=85mm]{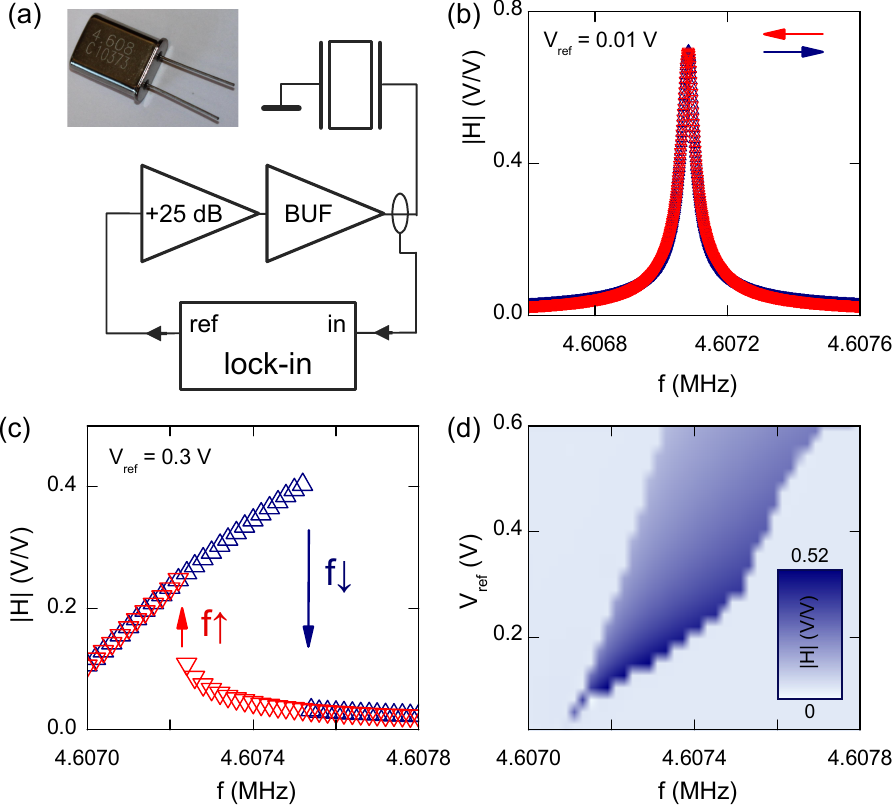}
\caption{(a) Schematic of the measurement circuit; inset: photograph of the quartz crystal. (b) Driven response (amplitude shown only) at the fundamental shear-mode driven at  $\mathrm{V_{ref}=0.01\,V}$. The arrows indicate the sweep direction. (c) Bistable response when driven at $\mathrm{0.3\,V}$. (d) Drive parameters that lead to a bistable shear-mode oscillation. The figure is obtained by superimposing forward and backward traces, such as the one shown in (c) (horizontal, fast axis), while varying the drive power (vertical, slow axis). Color scale: susceptibility of the quartz crystal.}
\end{figure}
\indent\indent The dynamic relaxation oscillator is implemented using a commercial AT-cut quartz crystal with a specified resonance frequency of 4.608 MHz at the fundamental thickness shear-mode. The crystal is driven by an rf-voltage generated by a lock-in amplifier, and its motion is transduced using a current probe~\cite{si_circuit} and detected by the lock-in, as is shown in Fig. 1(a). The experiments are performed at room temperature and in atmospheric pressure~\cite{environment}. Figure 1(b) shows the response of the weakly driven quartz crystal, which corresponds to a harmonic resonator, with a resonance frequency of $\mathrm{f_0=4.607\,MHz}$ and a mechanical ring-down time of $\mathrm{\tau_m = Q_0^{-1} \pi f_0 \approx 50\,ms}$.\\
\indent\indent When the driving voltage is increased, the response starts to deviate from a harmonic oscillator, as the resonance peak becomes non-symmetric. The resonance frequency shifts to a higher value: a signature of nonlinear behaviour with a positive higher order spring constant. The anharmonic behaviour of a quartz crystal has been studied previously~\cite{gagnepain81,vasiljevic87}, and in a recent quartz crystal microbalance (QCM) experiment, a nonlinear response was used to enhance the responsivity to an added mass~\cite{kirkendall13}.

\begin{figure}[h]
\includegraphics[width=85mm]{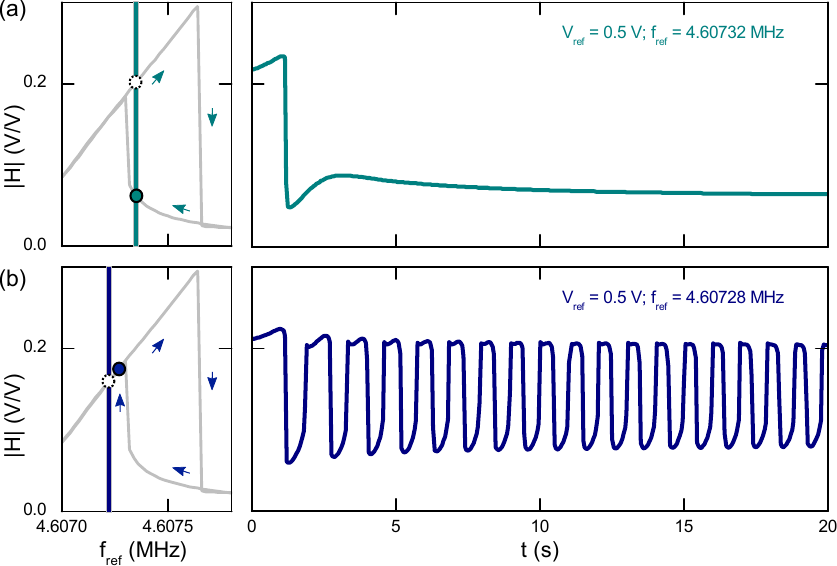}
\caption{(a) When the crystal is driven close to the bifurcation point inside the hysteresis regime, it decays to the low-amplitude state.  (b) When driven close to the bifurcation point outside the hysteresis regime, relaxation oscillations occur. }
\end{figure}

\indent\indent When the crystal is driven beyond the critical point, at $\mathrm{V_{ref}=0.02\,V}$ and $\mathrm{f_{ref}=4.6071\,MHz}$, the nonlinearity gives rise to a bifurcation and hysteresis occurs: two stable states co-exist, and the crystal oscillates either at a high or at a low amplitude.  Figure 1(c) shows a hysteretic frequency response of the quartz crystal. When the frequency is swept from a low to a high value, the crystal susceptibility follows the upper branch, and it oscillates at a high amplitude. For a reverse sweep, the low amplitude state is stable. Figure 1(d) shows the driving conditions that give rise to a bistable response. The figure is constructed by superimposing frequency response measurements taken in forward and in reverse direction (horizontal, fast axis), at a varying drive amplitude (vertical, slow axis). The hysteretic transitions between the two oscillating states, which occur at the bifurcation frequencies $\mathrm{f\downarrow}$ and $\mathrm{f\uparrow}$, form the switching element that is required for the relaxation oscillator.\\
\indent\indent\ The electrical impedance of a quartz crystal depends on the amplitude of the oscillation, and it is bistable when the amplitude is bistable, as in the regime visualized in Fig.1(d). We measured that $\mathrm{Z_{low}\approx 800\,\Omega}$ in the low state and $\mathrm{Z_{high}\approx 20\,\Omega}$ in the high state. Given the constant driving voltage, the dissipated power then depends on the oscillation state. Joule-heating of the crystal gives rise to a significant change in its mechanical properties. In particular,  the negative temperature dependence of the Young’s modulus of silicon oxide causes the mechanical spring constant, and thus the (nonlinear) resonance frequency, to decrease with temperature. As a result, the susceptibility of the crystal at the driving tone is increased, and this thermo-mechanical coupling presents the feedback that gives rise to relaxation oscillations.\\
\indent\indent Figure 2 shows time traces of the crystal amplitude when driven at  $\mathrm{V_{ref}=0.5\,V}$ at two different fixed frequencies. When driven close to $\mathrm{f\uparrow}$ at the high-amplitude branch, as shown in panel(a), the crystal temperature increases, causing the resonance frequency to decrease and the susceptibility to increase. The response then follows the upper branch as indicated by the arrows, until  $\mathrm{f\downarrow<f_{ref}}$, where the impedance becomes low. The crystal then cools and follows the lower branch until it reaches a steady low-amplitude state. Figure 2(b), right hand panel, shows a measured time trace of the crystal amplitude, with a single transition to the low state at $\mathrm{t=2\,s}$. A similar process occurs when the crystal is driven close to $\mathrm{f{\uparrow}}$ but outside the hysteretic regime, as in panel (b), but in this case the temperature further decreases, until $\mathrm{f\uparrow > f_{ref}}$ forces an upwards transition. Here, the process repeats, and the crystal enters a relaxation oscillation. Figure 2(b), right panel, shows a time trace of the oscillation, with a period of approximately $\mathrm{1\,s}$.\\
\indent\indent To investigate the relaxation oscillator in more detail, we measured time series of the crystal susceptibility close to the bifurcation point, $\mathrm{f\uparrow}$, for a range of driving voltages and frequencies. Figure 3(a), shows the frequency of the relaxation oscillation on the color scale, for a range of drive parameters. The bifurcation points, as measured in Fig.1(d), are indicated by the solid blue lines. In the dark blue area the frequency is zero and the relaxation oscillations are absent. Relaxation oscillations occur in a distinct regime close to $\mathrm{f\uparrow}$, at a frequency that can be adjusted over $\mathrm{f_{RO}=0..2.5\,Hz}$ by adjusting the drive parameters.\\

\begin{figure}[h]
\includegraphics[width=120mm]{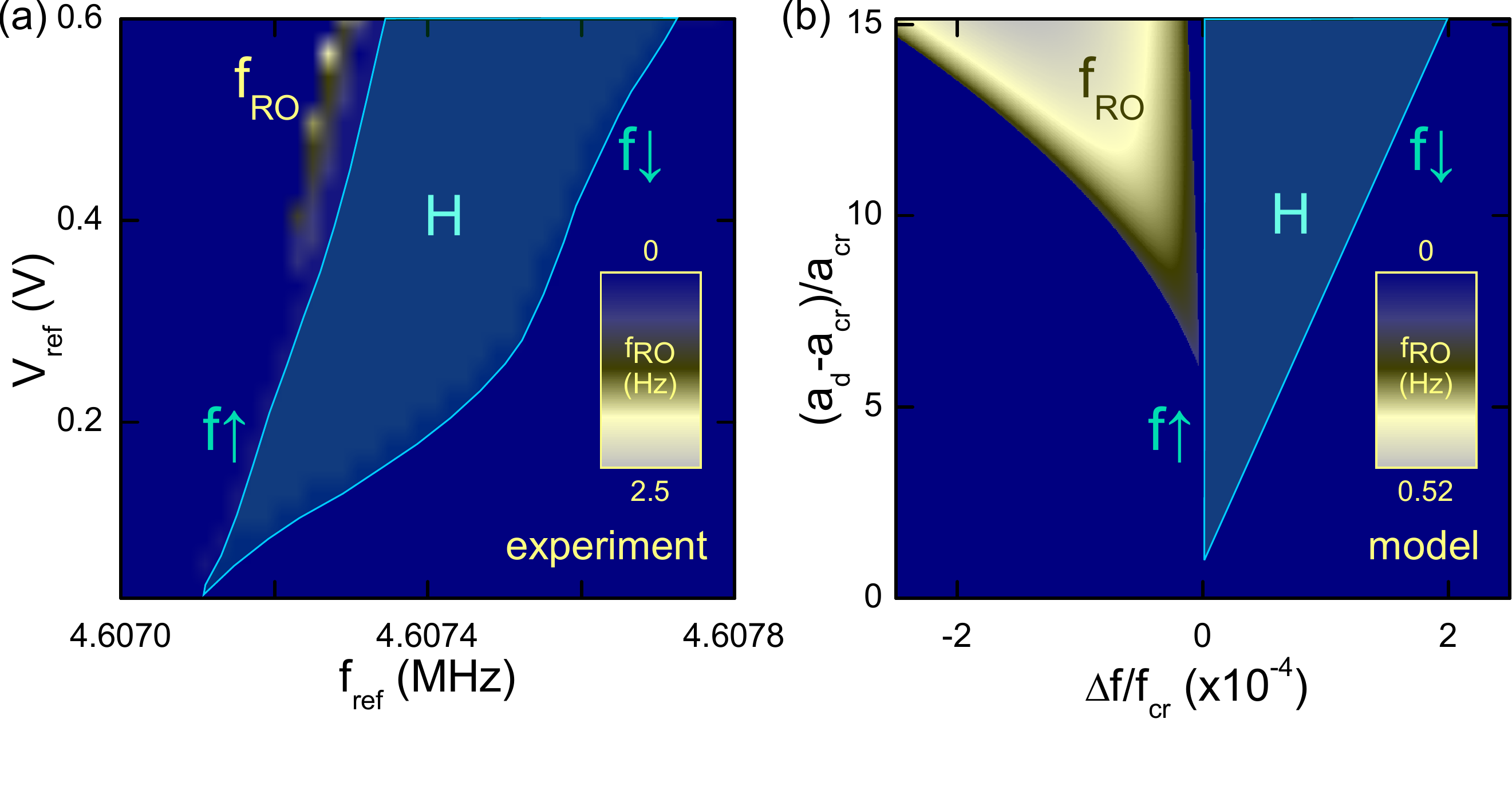}
\caption{Frequency of the relaxation oscillations (color scale) for a range of driving parameters, as obtained from experiment (a) and model (b). For each pixel, one time trace is measured from which the frequency is determined. In the model, the  drive strength, $\mathrm{a_d}$, and the frequency detuning, $\mathrm{\Delta f}$, are normalized to the critical drive strength and frequency, $\mathrm{a_{cr}}$ and $\mathrm{f_{cr}}$, respectively.}
\end{figure}

\indent\indent\ To corroborate the experimental results, a simple model is developed. While the power and frequency of the bifurcation points in a nonlinear resonator can be modelled following the analysis by Lifshitz and Cross~\cite{lifshitz08}, here we simplify the problem by taking only the slowest dynamics into account. We assume a constant susceptibility (dissipation) on the high and on the low branch of the hysteresis regime, and a linear dependence of the bifurcation frequencies on the drive strength, $\mathrm{f{\uparrow},f{\downarrow}\propto {V_{ref}}}$. We take a linear dependence of the bifurcation frequency on the crystal temperature, $\mathrm{f{\uparrow},f{\downarrow} \propto{T}}$ ~\cite{vittoz2010low}, and model the behaviour using a first order heat equation.\\ 
\indent\indent With the dissipation $\mathrm{P_{low}}$ in the lower branch and  $\mathrm{P_{high}}$ in the upper branch~\cite{genin78}, the temperature of the crystal is described as~\cite{hyde1971thermistors} 
\begin{equation}
\Delta T(t)=\frac{P}{K}-\left(\frac{P}{K}-\Delta T_i\right)e^{\frac{-t}{\tau}},
\end{equation}
where $t$ denotes time, $\tau$ is the thermal time constant of the crystal, and $\Delta T_i$ is the initial temperature difference with respect to room temperature. When $K$ is the thermal conductance between the crystal and the environment, $T_{\text{high}}=P_{\text{high}}/K$  and $T_{\text{low}}=P_{\text{low}}/K$  represent the two thermal equilibria to which the oscillator relaxes. By solving the heat equation and inserting $\mathrm{f\propto T}$, one obtains the thermal oscillation period:
\begin{equation}
f_{\mathrm{RO}}^{-1}=\tau \ln \left(\frac{f_{\mathrm{high}}-f{\uparrow}}{f_{\mathrm{high}}-f{\downarrow}}\right)+\tau \ln\left(\frac{f{\downarrow} - f_{\mathrm{low}}}{f{\uparrow}- f_{\mathrm{low}}}\right),
\end{equation}
\indent\indent The relaxation oscillation period is thus set by the difference between the thermal equilibrium frequencies $\mathrm{f_{low}}$ and $\mathrm{f_{high}}$, and the bifurcation points $\mathrm{f{\uparrow}}$ and $\mathrm{f{\downarrow}}$. Reducing the denominator in either of the terms in Eq.2 increases the cooling or heating time, and thus the oscillation period, as is observed in the experiment of Fig. 3(a). Equation (2) was solved for a range of excitation frequency and amplitudes with  $\tau=1$, $K = 1$, $P_{\text{high}}/P_{\text{low}}$ = 40, and the frequency of the relaxation oscillation, $\mathrm{f_{RO}}$, is plotted in Fig.~3(b). As in the experiment, relaxation oscillations occur for $\mathrm{f_{ref}<f\uparrow}$. Moreover, the frequency of the relaxation oscillation increases with the detuning and it exhibits a maximum. The model could be refined to capture the experimentally observed features in more detail. For instance, the frequency-dependence of the temperature in the high and in the low amplitude state could be taken into account, and the kink in the $\mathrm{f\downarrow}$ branch could be modeled by accounting for higher-order nonlinearities or multiple (internal) resonance modes~\cite{kirkendall13}.\\
\indent\indent The dissipated power can be estimated from the impedances in the high and low states, and we calculate that $\mathrm{P_{high}\approx 2.5\,W}$ and  $\mathrm{P_{low}\approx 0.1\,W}$, compared to a dissipation of ~1 mW in the linear regime of Fig.1(b). It is interesting to obtain an indication of the temperature range during the relaxation cycle. To this end, the crystal was removed from its package, and its temperature was measured using an infrared detector~\cite{tsensor}, facing the crystal at a distance of 2 mm. Figure 4 shows the temperature of the crystal during the oscillation. When driven at $\mathrm{f_{ref}=4.60727\,MHz}$ and $\mathrm{V_{ref}=0.55\,V}$ in an ambient temperature of $\mathrm{20\,\degree C}$, the mean temperature is $\mathrm{35\,\degree C}$, and the peak-to-peak value is $\mathrm{7\,\degree C}$. The frequency of the relaxation oscillation is slightly shifted compared to a packaged crystal due to a different thermal conductance, and the oscillation is less stable due to the exposure to the environment. \\
\indent\indent Relaxation oscillators are very sensitive to their environment, as is indicated by the frequency fluctuations that can be observed in the measurements of Fig.2(b) and Fig.4, in which no special precautions were taken for stabilization. This sensitivity could be exploited in detectors~\cite{genin78}: the relaxation rates respond, besides to mass, to changes in the viscosity and the thermal conductivity of the environment, enabling application as a viscosity or a pressure sensor. Since the output of the oscillator resembles a frequency- and pulse-width modulated binary signal, the interface to a digital circuit could be simplified. The quartz crystal may also be used as an experimental platform for more fundamental studies on dynamic relaxation oscillations, such as the complex but slow behaviour that occurs in electrochemical systems~\cite{baba08}. Here, the mechanical device presents interesting dynamics on an experimentally convenient time scale. Finally, we note that the thermal time constant scales linearly with dimension (volume-to-surface ratio), and that for micrometer-sized devices similar processes could occur at frequencies in the kHz range.\\
 \begin{figure}[h]
\includegraphics[width=80mm]{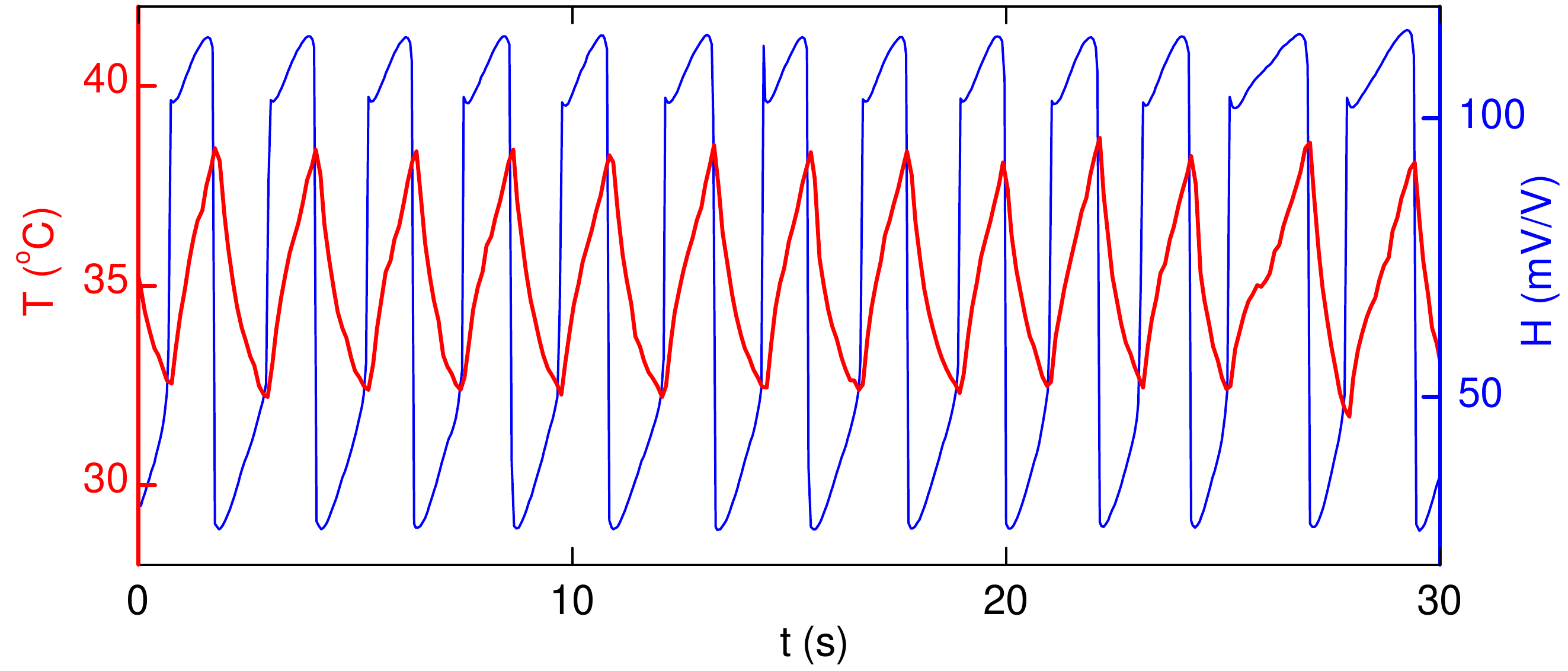}
\caption{Temperature of the quartz crystal (red trace, left axis) during the relaxation cycle (blue trace, right axis), measured after removing the crystal from the package. The crystal is driven at $\mathrm{f_{ref}=4.60727\,MHz}$, and the driving amplitude is $\mathrm{V_{ref}=0.55\,V}$.}
\end{figure}
\indent\indent In conclusion, we demonstrate dynamic relaxation oscillations in a strongly driven quartz crystal. The amplitude of the crystal oscillates at a frequency that is determined by the thermal relaxation time of the crystal, six orders of magnitude lower the fundamental harmonic oscillation frequency of the crystal. The dynamic behaviour is captured by a model that takes into account the slow dynamics. The frequency of the relaxation oscillation is very sensitive to the susceptibility of the crystal, which depends on its mechanical properties. This feature may be deployed in a low-frequency readout scheme for QCM-based sensors.

\indent\indent\ The authors acknowledge financial support from the European Union's Seventh Framework Programme (FP7) under grant agreement $\mathrm{n{\circ}~318287}$ (project LANDAUER), and an STW Take-off grant.

%\bibliographystyle{apsrev}
%\bibliography{QCRO}

\begin{thebibliography}{18}
\expandafter\ifx\csname natexlab\endcsname\relax\def\natexlab#1{#1}\fi
\expandafter\ifx\csname bibnamefont\endcsname\relax
  \def\bibnamefont#1{#1}\fi
\expandafter\ifx\csname bibfnamefont\endcsname\relax
  \def\bibfnamefont#1{#1}\fi
\expandafter\ifx\csname citenamefont\endcsname\relax
  \def\citenamefont#1{#1}\fi
\expandafter\ifx\csname url\endcsname\relax
  \def\url#1{\texttt{#1}}\fi
\expandafter\ifx\csname urlprefix\endcsname\relax\def\urlprefix{URL }\fi
\providecommand{\bibinfo}[2]{#2}
\providecommand{\eprint}[2][]{\url{#2}}

\bibitem[{\citenamefont{Jenkins}(2013)}]{jenkins13}
\bibinfo{author}{\bibfnamefont{A.}~\bibnamefont{Jenkins}},
  \bibinfo{journal}{Phys Rep} \textbf{\bibinfo{volume}{525}},
  \bibinfo{pages}{167} (\bibinfo{year}{2013}).

\bibitem[{\citenamefont{Abraham and Bloch}(1919)}]{abraham19}
\bibinfo{author}{\bibfnamefont{H.}~\bibnamefont{Abraham}} \bibnamefont{and}
  \bibinfo{author}{\bibfnamefont{E.}~\bibnamefont{Bloch}}, \bibinfo{journal}{J
  Phys Theor Appl} \textbf{\bibinfo{volume}{9}}, \bibinfo{pages}{211}
  (\bibinfo{year}{1919}).

\bibitem[{\citenamefont{Van~der Pol and Van~der Mark}(1928)}]{vdpol28}
\bibinfo{author}{\bibfnamefont{B.}~\bibnamefont{Van~der Pol}} \bibnamefont{and}
  \bibinfo{author}{\bibfnamefont{J.}~\bibnamefont{Van~der Mark}},
  \bibinfo{journal}{The London, Edinburgh, and Dublin Philosophical Magazine
  and Journal of Science} \textbf{\bibinfo{volume}{6}}, \bibinfo{pages}{763}
  (\bibinfo{year}{1928}).

\bibitem[{\citenamefont{Pearson and Anson}(1921)}]{pearson21}
\bibinfo{author}{\bibfnamefont{S.~O.} \bibnamefont{Pearson}} \bibnamefont{and}
  \bibinfo{author}{\bibfnamefont{H.~S.~G.} \bibnamefont{Anson}},
  \bibinfo{journal}{Proc Phys Soc London} \textbf{\bibinfo{volume}{34}},
  \bibinfo{pages}{175} (\bibinfo{year}{1921}).

\bibitem[{\citenamefont{Bienstman et~al.}(1998)\citenamefont{Bienstman,
  Vandewalle, and Puers}}]{bienstman98}
\bibinfo{author}{\bibfnamefont{J.}~\bibnamefont{Bienstman}},
  \bibinfo{author}{\bibfnamefont{J.}~\bibnamefont{Vandewalle}},
  \bibnamefont{and} \bibinfo{author}{\bibfnamefont{R.}~\bibnamefont{Puers}},
  \bibinfo{journal}{Sensor Actuat A: Phys} \textbf{\bibinfo{volume}{66}},
  \bibinfo{pages}{40} (\bibinfo{year}{1998}).

\bibitem[{\citenamefont{Lazarus et~al.}(2010)\citenamefont{Lazarus, Barois,
  Perisanu, Poncharal, Manneville, De~Langre, Purcell, Vincent, and
  Ayari}}]{lazarus10}
\bibinfo{author}{\bibfnamefont{A.}~\bibnamefont{Lazarus}},
  \bibinfo{author}{\bibfnamefont{T.}~\bibnamefont{Barois}},
  \bibinfo{author}{\bibfnamefont{S.}~\bibnamefont{Perisanu}},
  \bibinfo{author}{\bibfnamefont{P.}~\bibnamefont{Poncharal}},
  \bibinfo{author}{\bibfnamefont{P.}~\bibnamefont{Manneville}},
  \bibinfo{author}{\bibfnamefont{E.}~\bibnamefont{De~Langre}},
  \bibinfo{author}{\bibfnamefont{S.~T.} \bibnamefont{Purcell}},
  \bibinfo{author}{\bibfnamefont{P.}~\bibnamefont{Vincent}}, \bibnamefont{and}
  \bibinfo{author}{\bibfnamefont{A.}~\bibnamefont{Ayari}},
  \bibinfo{journal}{Appl Phys Lett} \textbf{\bibinfo{volume}{96}},
  \bibinfo{pages}{193114} (\bibinfo{year}{2010}).

\bibitem[{\citenamefont{Regan et~al.}(2005)\citenamefont{Regan, Aloni, Jensen,
  and Zettl}}]{zettl05}
\bibinfo{author}{\bibfnamefont{B.~C.} \bibnamefont{Regan}},
  \bibinfo{author}{\bibfnamefont{S.}~\bibnamefont{Aloni}},
  \bibinfo{author}{\bibfnamefont{K.}~\bibnamefont{Jensen}}, \bibnamefont{and}
  \bibinfo{author}{\bibfnamefont{A.}~\bibnamefont{Zettl}},
  \bibinfo{journal}{Appl Phys Lett} \textbf{\bibinfo{volume}{86}},
  \bibinfo{pages}{123119} (\bibinfo{year}{2005}).

\bibitem[{si_()}]{si_circuit}
\emph{\bibinfo{title}{\emph{See supplemental material at
  \url{ftp://ftp.aip.org/epaps/appl_phys_lett/E-APPLAB-107-038533/} for a
  schematic of the measurement circuit.}}}

\bibitem[{env()}]{environment}
\emph{\bibinfo{title}{\emph{The crystal is mounted in a standard hermetically
  sealed $\mathrm{HC-49/U}$-type nitrogen-filled metal package.}}}

\bibitem[{\citenamefont{Gagnepain}(1981)}]{gagnepain81}
\bibinfo{author}{\bibfnamefont{J.~J.} \bibnamefont{Gagnepain}}, in
  \emph{\bibinfo{booktitle}{Thirty Fifth Annual Frequency Control Symposium.
  1981}} (\bibinfo{year}{1981}), pp. \bibinfo{pages}{14--30}.

\bibitem[{\citenamefont{Vasiljevic}(1987)}]{vasiljevic87}
\bibinfo{author}{\bibfnamefont{D.~M.} \bibnamefont{Vasiljevic}},
  \bibinfo{journal}{IEEE Trans Circuits and Systems}
  \textbf{\bibinfo{volume}{34}}, \bibinfo{pages}{897} (\bibinfo{year}{1987}).

\bibitem[{\citenamefont{Kirkendall et~al.}(2013)\citenamefont{Kirkendall,
  Howard, and Kwon}}]{kirkendall13}
\bibinfo{author}{\bibfnamefont{C.~R.} \bibnamefont{Kirkendall}},
  \bibinfo{author}{\bibfnamefont{D.~J.} \bibnamefont{Howard}},
  \bibnamefont{and} \bibinfo{author}{\bibfnamefont{J.~W.} \bibnamefont{Kwon}},
  \bibinfo{journal}{Appl Phys Lett} \textbf{\bibinfo{volume}{103}},
  \bibinfo{pages}{223502} (\bibinfo{year}{2013}).

\bibitem[{\citenamefont{Lifshitz and Cross}(2008)}]{lifshitz08}
\bibinfo{author}{\bibfnamefont{R.}~\bibnamefont{Lifshitz}} \bibnamefont{and}
  \bibinfo{author}{\bibfnamefont{M.~C.} \bibnamefont{Cross}},
  \bibinfo{journal}{Reviews of nonlinear dynamics and complexity}
  \textbf{\bibinfo{volume}{1}}, \bibinfo{pages}{1} (\bibinfo{year}{2008}).

\bibitem[{\citenamefont{Vittoz}(2010)}]{vittoz2010low}
\bibinfo{author}{\bibfnamefont{E.~A.} \bibnamefont{Vittoz}},
  \emph{\bibinfo{title}{Low-Power Crystal and MEMS Oscillators: The Experience
  of Watch Developments}}, Integrated Circuits and Systems
  (\bibinfo{publisher}{Springer}, \bibinfo{year}{2010}).

\bibitem[{\citenamefont{Genin and Brezel}(1978)}]{genin78}
\bibinfo{author}{\bibfnamefont{R.}~\bibnamefont{Genin}} \bibnamefont{and}
  \bibinfo{author}{\bibfnamefont{P.}~\bibnamefont{Brezel}},
  \bibinfo{journal}{Int J Electr Theor Exp} \textbf{\bibinfo{volume}{45}},
  \bibinfo{pages}{97} (\bibinfo{year}{1978}).

\bibitem[{\citenamefont{Hyde}(1971)}]{hyde1971thermistors}
\bibinfo{author}{\bibfnamefont{F.~J.} \bibnamefont{Hyde}},
  \emph{\bibinfo{title}{Thermistors}} (\bibinfo{publisher}{Iliffe},
  \bibinfo{year}{1971}).

\bibitem[{tse()}]{tsensor}
\emph{\bibinfo{title}{\emph{Melexis MLX90614 infrared thermometer}}}.

\bibitem[{\citenamefont{Baba and Krischer}(2008)}]{baba08}
\bibinfo{author}{\bibfnamefont{N.}~\bibnamefont{Baba}} \bibnamefont{and}
  \bibinfo{author}{\bibfnamefont{K.}~\bibnamefont{Krischer}},
  \bibinfo{journal}{Chaos: An Interdisciplinary Journal of Nonlinear Science}
  \textbf{\bibinfo{volume}{18}}, \bibinfo{pages}{015103}
  (\bibinfo{year}{2008}).

\end{thebibliography}

\end{document}